# Network model and analysis of the spread of Covid-19 with social distancing

Parul Maheshwari[1], Reka Albert[1,2]

1. Department of Physics, The Pennsylvania State University, University Park, PA-16802
2. Biology Department, The Pennsylvania State University, University Park, PA-16802

**Abstract**

The first mitigation response to the Covid-19 pandemic was to limit person-to-person interaction as much as possible. This was implemented by the temporary closing of many workplaces and people were required to follow social distancing. Networks are a great way to represent interactions among people and the temporary severing of these interactions. Here, we present a network model of human-human interactions that could be mediators of disease spread. The nodes of this network are individuals and different types of edges denote family cliques, workplace interactions, interactions arising from essential needs, and social interactions. Each individual can be in one of four states: susceptible, infected, immune, and dead. The network and the disease parameters are informed by the existing literature on Covid-19. Using this model, we simulate the spread of an infectious disease in the presence of various mitigation scenarios. For example, lockdown is implemented by deleting edges that denote non-essential interactions. We validate the simulation results with the real data by matching the basic and effective reproduction numbers during different phases of the spread. We also simulate different possibilities of the slow lifting of the lockdown by varying the transmission rate as facilities are slowly opened but people follow prevention measures like wearing masks etc. We make predictions on the probability and intensity of a second wave of infection in each of these scenarios.

### I. Introduction

The spread of the novel SARS-CoV-2 virus has posed a new challenge to the scientific community and society at large on mitigating the spread of a viral infection. The lack of vaccines or therapies for the virus calls for non-pharmaceutical intervention in the course of disease spread. As a result, the immediate response from most of the world was to limit the contact among people in the population by requiring them to stay at home. The person-level effect of these measures is to maintain physical distance with everyone except co-habiting family members and reduce social interactions. The expected effect of this distancing is the reduction in the spread of the virus, widely referred to as "flattening the curve". While the real-time data shows a reduction in the rate of spread in situations where social distancing was strictly followed [1], analysis to predict the exact effect of the mitigation measures will be useful to understand the possibility of a second wave of infection and choose an optimal strategy to minimize the effect of the pandemic while minimizing the economic impact of the restrictions. Different epidemiological modeling approaches modeled this reduced contact as modified parameters of the rate of spread of the infection. Most of these epidemiological models are compartmental models that assume a fraction of the population as either susceptible (S), infected (I), or recovered/dead (R) and model the rate of change of these continuous variables (S, I, and R) [2-4]. There are also network-based epidemiological models that represent human-human interactions as edges and model the spread of the disease by infecting a fraction of the neighbors of an infected node [5-9]. The fraction infected is varied as the population starts to follow social distancing measures.

Here, we present a network model where the effect of lockdown or social distancing leads to a modification in the network structure. We construct a network for a given population size so that it comprises of two different kinds of edges, one that is deleted in lockdown and other that continues to exist during lockdown. These two kinds of edges denote the two different kinds of interactions we have in our daily lives. The edges that are preserved during lockdown denote interactions with family members and essential service providers and the edges that are deleted during lockdown denote other "non-essential" interactions like workplace interactions, socializing at events and club meetings, etc. We find that this way to simulate the social distancing in the population gives promising results in terms of reduction of the spread of the virus. We then explore different periods of lockdown with different strategies of phasing out of the lockdown and find interesting insight on the possibility of a second wave as a result.

## II. Methods

### A. Construction of the network

We construct networks of 10,000 nodes. Each node corresponds to a person and an edge between nodes corresponds to a non-zero probability of disease transmission between the two persons. Each person is assumed to be a part of a family. The network nodes do not include anyone under the age of 18 on the assumption that the disease has little effect on children. Family sizes range from 1 (with frequency of

30%), 2 (with frequency of 35%), 3 (18%) to 4 (17% frequency); this data was obtained and approximated from [10]. Each family is a clique in the network. We assume that 70% of the population is made up by working individuals. We approximate this number from the Wikipedia page on Employment-to-Population ratio [11]. We assume that 20% of the working population are essential workers. We construct a scale-free network [12] with a minimum degree of 2 and degree distribution $P(k) \sim k^{-3}$ among the non-essential working population. We assume that network among the essential working population is random [13] with average degree 25. Additionally, there are edges that represent interaction between the population and essential workers. This network covers the interactions that happen due to essential activities like grocery shopping, healthcare, etc. This is a random network with the probability of an edge to an essential worker set to 0.2. Finally, some edges are added to the network to cover social interaction outside of families and workplaces in the population. This is a random network with average degree of ~1.3.

### B. Dynamics of the spread

Each 10,000-node network is initialized with 5 infected nodes, selected randomly, and the rest of the nodes are susceptible. Each infected node can infect its neighbors (i.e. nodes that are connected to the infected node by an edge) with a certain transmission probability. The transmission probability depends on whether the infected person is symptomatic or asymptomatic. Since symptomatic people are likely visit the doctor, and then they isolate themselves at a hospital or at home, we assume that their transmission probability to essential workers (i.e. health workers) is 0.05 and their transmission probability to the general population is 0.005. We assume the transmission probability for asymptomatic infected persons is 0.05. Additionally, we assume that 80% of the infected nodes are asymptomatic while the other 20% are symptomatic. From the data on COVID-19 death rates, we assume a 5% mortality among the infected nodes. These 5% form a subset of the symptomatic patients. After the other 95% recover, 85% become immune to the disease while the remaining 10% become susceptible again. The number of days it takes for a person to recover from the infection varies from 1 day to 35 days, following a normal distribution with a mean of 17 days and standard deviation of 4.8 days. Similarly, the number of days after which a patient loses their life to the disease varies from 5 days to 24 days according to a normal distribution centered at 14 days and standard deviation of 3.2 days. In the unmitigated situation, an infected node can spread the infection to any of its neighbors. During a lockdown (i.e. period when the population is following social distancing and stay-at-home order), only certain edges of the network are retained for the spread of the disease. These edges correspond to family interactions, interactions among essential workers, and interactions between the population and essential workers. We run this simulation for many graphs, and for each graph, we run the spread process multiple times. In total, we take an average over 50 runs and report the results.

### C. Reproduction number of the disease

The reproduction number is a measure of the expected number of cases directly generated by one infected case. The basic reproduction number is this measure in a population where everyone is susceptible, and the effective reproduction number ($R_t$) is for a population where some fraction of the population is immune. In this network modeling, we calculate the reproduction number by counting the number of people an infected person spreads during the time they were sick. For example, consider node A who gets infected on day 1 and node infects 2 persons on day 2, 1 person on day 3, 2 persons on day 4. Node A recovers on day 5 and cannot infect any more persons. So, node A infected 5 persons in total and hence the reproduction number for node A is 5. To calculate the disease's effective reproduction number for day 1, we find the reproduction number for each node that got infected on day 1 and take the average. Hence, the reproduction number of a day is the average reproduction number for every person that caught the infection on that day.

The code for the network generation, spread of the disease, and calculation of the reproduction number of the disease, is freely available on GitHub: https://github.com/parulm/spread_lockdown.

## III. Results

### A. Network model

The constructed network is a combination of scale-free and random networks with multiple cliques of sizes 1-4. Hence, in the resulting network, almost every node is a part of a clique, most of the nodes have a low degree (<10) but there also are some hubs (with degree >100). As illustration we show a 100-node network constructed by this method (Figure 1). This sample network has average degree of ~5 with maximum degree of 21. It has 10 connected components, the largest of which contains 86 nodes.

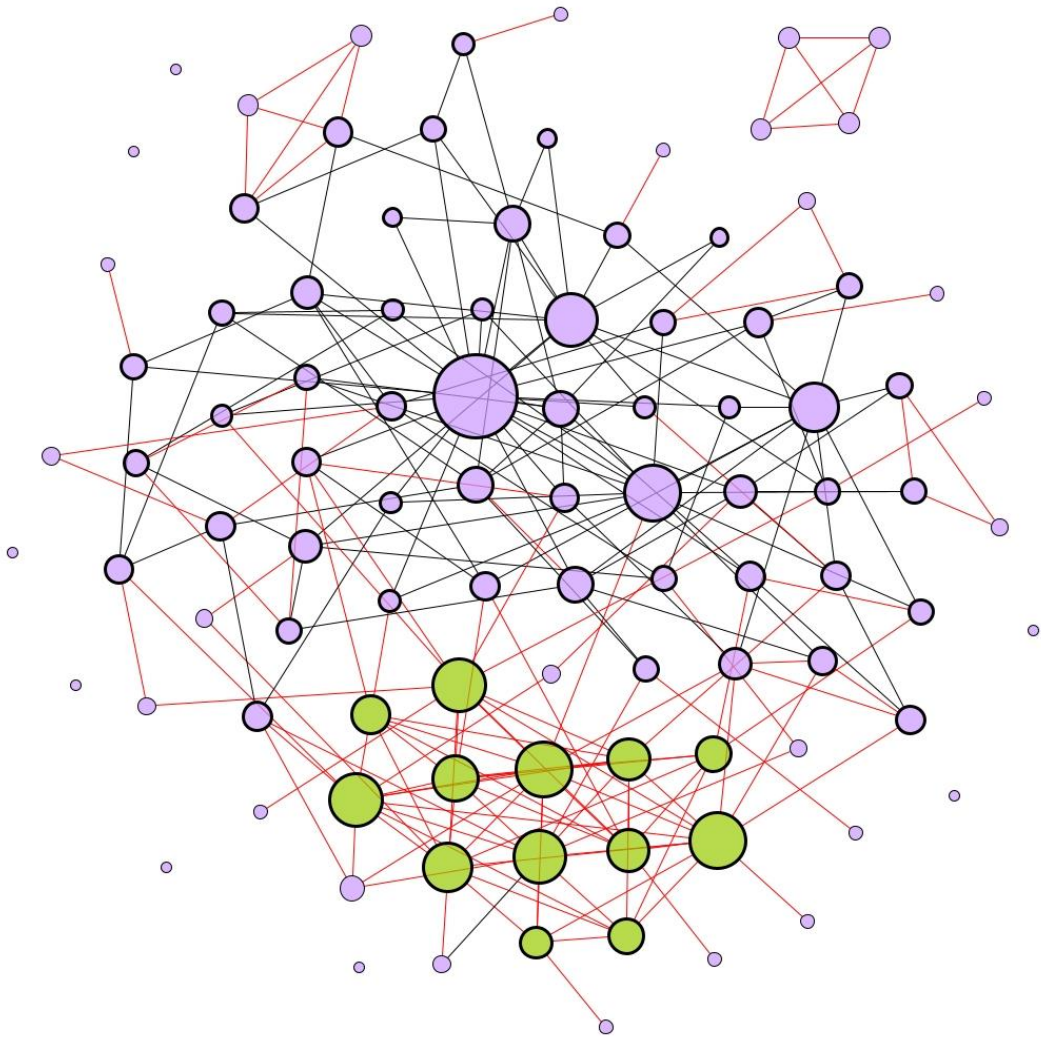

Figure 1. Illustration of the result of the network generation method on a population of 100 individuals. The edges represent interactions within families, workplaces, and social interaction. The red color edges are the ones that are preserved in the lockdown while the black color edges are deleted. Node size is directly proportional to the degree of the node. The nodes that have a thick border denote the people who are working. Nodes in green color denote essential workers.

The centrality measures of a network generated using this method vary as a result of the randomness and stochasticity in the generation process, however the network still has some common features. The degree distribution of the network shows two peaks, one at low degrees (6-7) and the other centered around 25, the average degree of the essential workplace network; it also has a significantly long tail, see Figure 2. The average degree of a 10,000-node network is ~10 and the maximum degree is 184. The network has nearly 50 connected components, the largest of which contains with 9950 nodes.

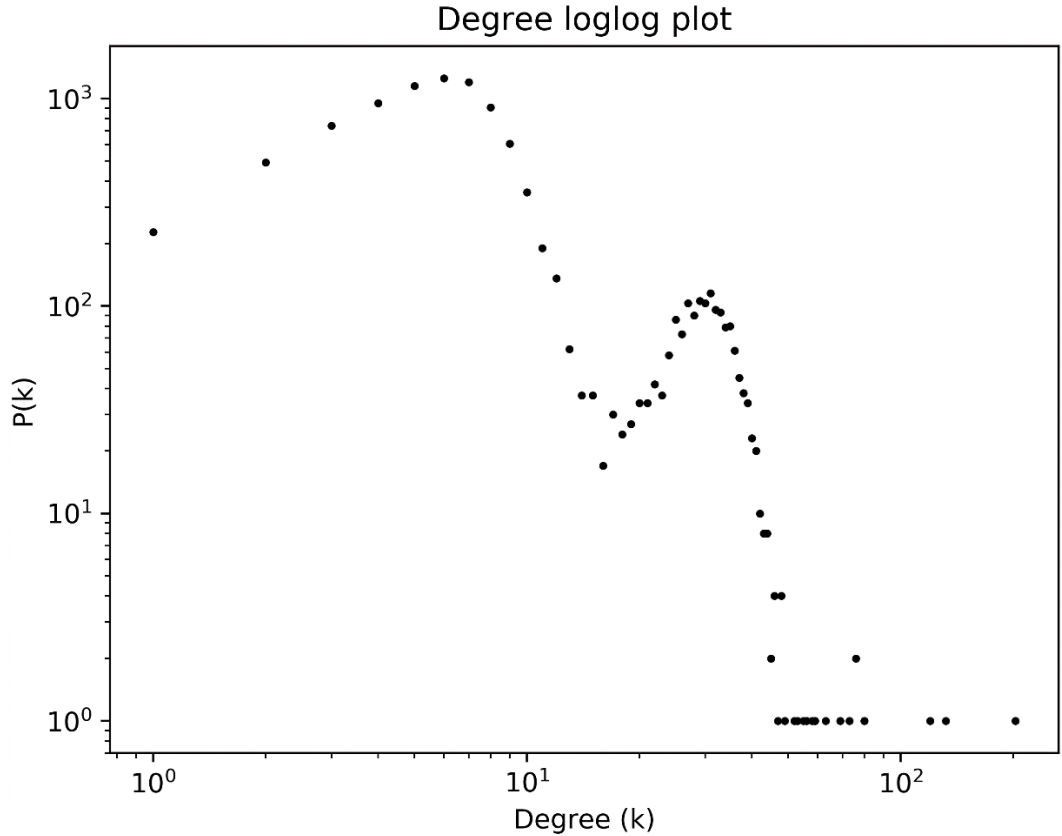

Figure 2. Degree distribution of a network generated with 10,000 nodes.

### B. Simulations of spread

In the absence of mitigation measures, the disease either dies off or spreads in the population. The likelihood of the epidemic taking off is 0.73. In the simulations where the epidemic takes off, the effective reproduction number starts at a high value (larger than 4), as expected from the theory of epidemic spreading (see Figure 3). In this case, the disease completes its course in ~80 days, infecting almost everyone. The peak of infection is around day 30, when ~45% of the population is infected at the same time. This disease time course leads to ~400 deaths, with very few deaths after day 80. The reproduction number peaks at ~7 around day 3 and goes to below 1 around day 20. Nearly 60% of the population becomes immune to the disease but over 30% of the population is susceptible and could be affected by a second wave of the disease. Since herd immunity needs that at least 70%-80% of the population be immune [14], letting the disease run its course unmitigated is not expected to benefit society in any way. We note that the number of deaths is only valid if the assumption of a constant death rate (of 5%) holds. It is likely that after day 10, when the hospitals start getting overwhelmed, the patients do not get all the medical attention they need, thus the death rate may increase.

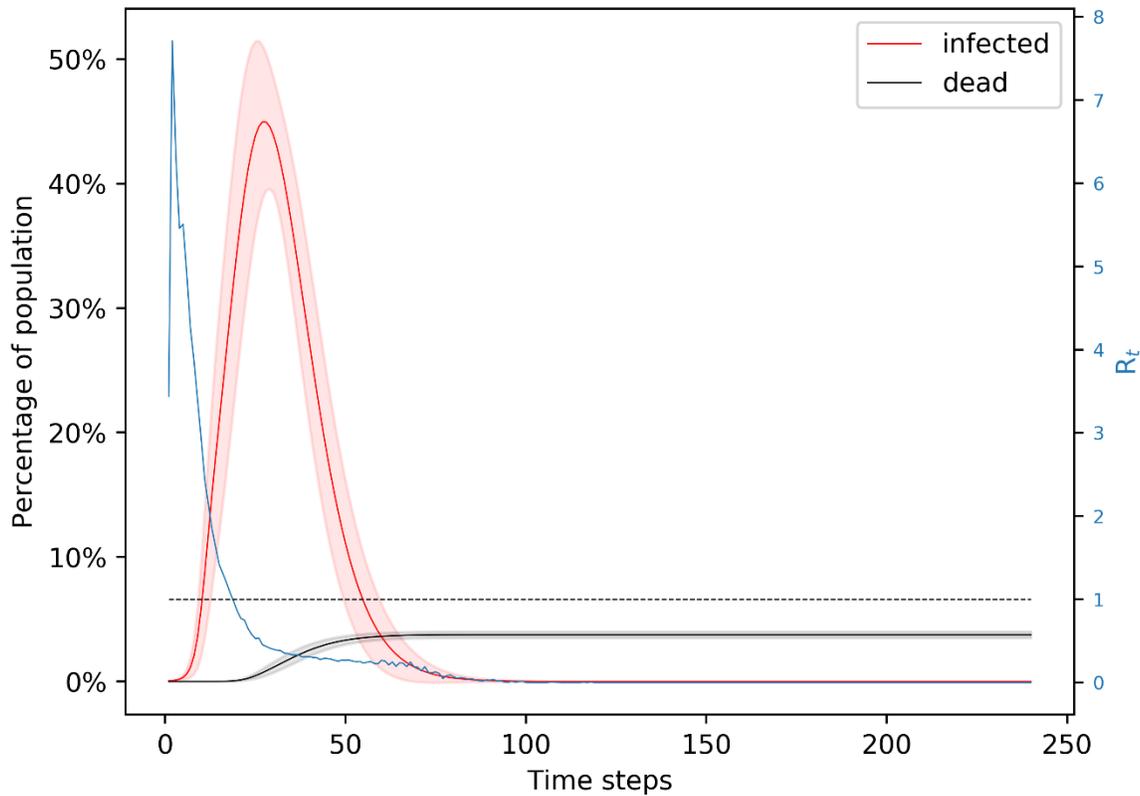

Figure 3. Unmitigated spread of the disease in the network model. The plot shows the number of nodes in the network that are infected and dead at a given time in red and black colors respectively. The reproduction number is shown in blue with the scale shown on the right y-axis. The x-axis gives the timesteps. The shaded area around the line plots is marked by the standard deviation obtained by running the simulation ~1000 times.

### A. Lockdown is fixed to 3 months

If a lockdown is implemented for 3 months with varying starting points, we find that it almost completely avoids the possibility of a second outbreak. We try the lockdown periods for day 5 to day 95, day 10 to day 100, and day 15 to day 105. All these cases lead to more than 150 deaths, with the number of deaths being higher if the lockdown starts later. Shown in Figure 4A, when the lockdown is from day 5 to day 95, the infection peaks at ~20% of the population. In nearly 1% of the simulations, there is a second wave of infection which peaks at less than 5% of the population infected around the same time. This wave of infection peaks nearly 50 days after the end of the lockdown. After 240 days, this scenario leads to a total of 150 deaths which is 1.5% of the population. Shown in Figure 4B, when the lockdown is from day 10 to day 100, the infection peaks to a little over 20% of the population. This scenario leads to nearly 0.7% of the simulations showing a small second wave of infection that affect a very small fraction of the population. After 240 days, this scenario leads to a total of 155 deaths. Figure 4C shows the scenario when lockdown is from day 15 to day 105. Nearly 100% of the simulations in this case lead to only the first wave of infection which peaks to ~25% of the population. After 240 days, this leads to ~175 deaths.

A high peak of infection implies that the medical facilities will be overwhelmed which is likely to cause a surge in the number of deaths. While these simulations show that a 90-day lockdown is beneficial for mitigating a widespread effect of the pandemic, they also show that the earlier a lockdown starts, the less the number of deaths and the less pressure on medical facilities. It is also interesting to note that an earlier lockdown leads to a slightly higher possibility of a second wave of infection since a smaller fraction of the population becomes immune. However, despite two waves of infection, the number of deaths is still lesser if the lockdown starts earlier. In all of these cases, the reproduction number peaks at ~6 around day 3-4, denoting the high infection spread in the following few days and it goes to below 1 around day 20. These results show that a 90-day lockdown starting the earliest possible would be the most efficient way to mitigate the effects of the disease through social distancing measures. However, due to the adverse economic effects of a lockdown, it is useful to find equally effective strategies that are shorter than a 90-day period. Hence, we next explore the effect of a 60-day lockdown.

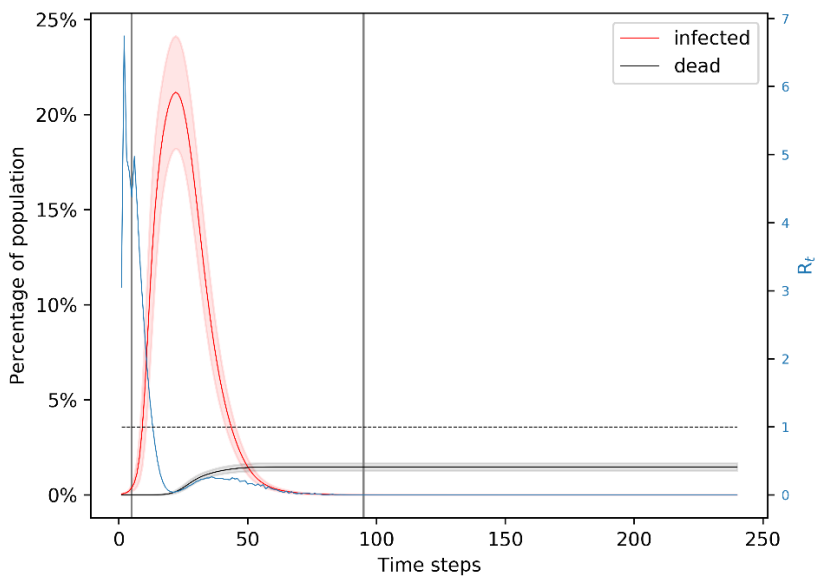

Figure 4A. Lockdown is from day 5 to day 95.

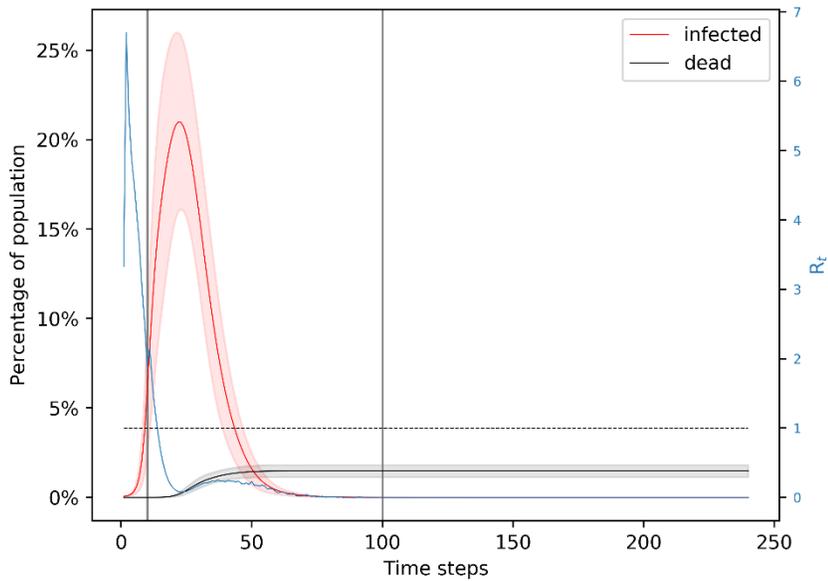

Figure 4B. Lockdown is from day 10 to day 100.

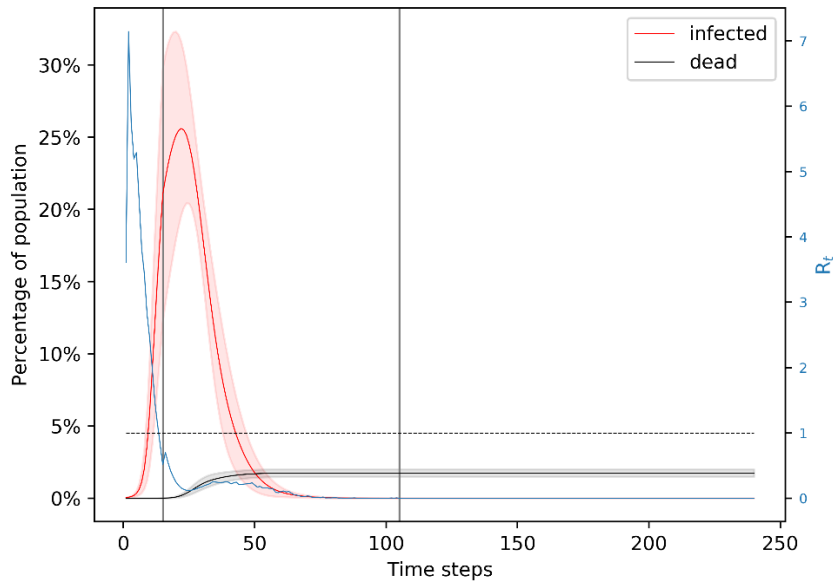

Figure 4C. Lockdown is from day 15 to day 105.

### B. Lockdown is fixed to 2 months

When the lockdown is fixed to 60 days and the starting time is varied, we find that a second wave of infection becomes more likely than in the 90-day lockdown case. Like the 90-day lockdown case, we explore 60-day lockdown for starting times of day 5, day 10 and day 15. The simulation of the lockdown from day 5 to day 65 shows that there is a second wave of infection in ~40% of the cases (Figure 5A) and only one wave of infection in 24% of the cases (Figure 5B); the remaining 36% of cases show no spread. The first wave of the infection in response to a day 5 to day 65 lockdown peaks around day 20 and infects nearly 20% of the population at the same time – this can be seen in both Figures 5A and 5B. As shown in Figure 5A, the second wave of infection peaks around day 110 and it is highly varying across simulations. On average, the peak of this wave infects another 10% of the population at the same time. Simulations with two waves of infection lead to a total of ~300 deaths which is 3% of the population. Simulations with only one wave of infection lead to a total of ~150 deaths. On average, the day 5 to day 65 lockdown scenario leads to ~250 deaths after 240 days.

We next explore the case when the lockdown is from day 10 to day 70. This results in ~33% of simulations showing a second wave of infection (Figure 6A), ~39% of simulations show only one wave of infection (Figure 6B) and the remaining 28% of the simulations show no spread. The second wave of infection peaks around day 110 and infects another 10% of the population at the same time. Two waves of infection lead to ~310 deaths and only one wave of infection leads to ~155 deaths. On average, the day 10 to day 70 lockdown scenario leads to ~220 deaths after 240 days. The day 15 to day 75 lockdown scenario leads to two waves of infection in ~16% of the simulations (Figure 7A), only one wave of infection in ~58% of the simulations (Figure 7B) and no spread in the remaining 26% of the simulations. In this case, the second wave of infection peaks around day 120 and infects another ~7% of the population at the same time. Two waves of infections lead to ~300 deaths and only one wave of infection leads to ~180 deaths. On average, the day 15 to day 75 lockdown scenario leads to 200 deaths.

In all these results, there is at least one peak for the effective reproduction number and this first peak is ~6 around day 3-4. In the results that show a second wave of infection, $R_t$ goes to below 1 during the lockdown but increases again after the end of lockdown; this peak is ~20 days before the peak in the number of infectious people. From these simulations, we can see that the probability of a second wave of infection is lower if the lockdown starts later. However, the first wave of the infections gets increasingly worse if the lockdown starts later. It makes sense to start the lockdown earlier to minimize the effects of the first lockdown, but we need other mechanisms to lower the effect of the second wave of infection. This can be done in various ways. For example, impose a second lockdown in response to the second wave of infection and so on until a consequent wave of infection is too small to affect the population adversely. In the next section, we consider another similar scenario where after the end of the first lockdown, the population resumes normal life gradually.

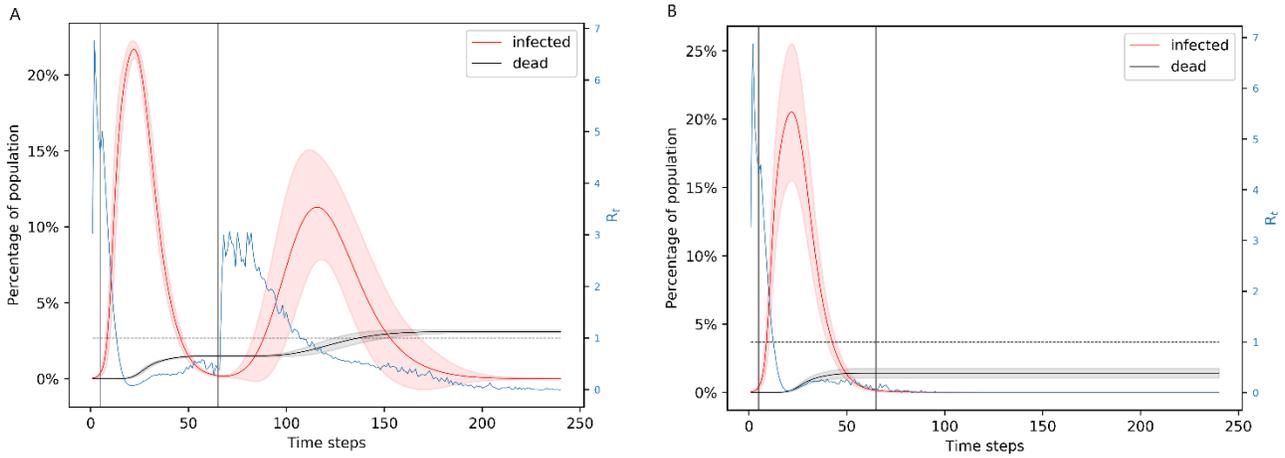

Figure 5**A**. Representation of the ~40% of simulation results where there is a second wave of infection when the lockdown is from day 5 to day 65. The effective reproduction number increases dramatically after the end of the lockdown, foreshadowing the second peak of the infection. **B**. Representation of the ~24% simulation results where there is no second wave of infection for the lockdown period of day 5 to day 65. The effective reproduction number stays below 1 following the lifting of the lockdown.

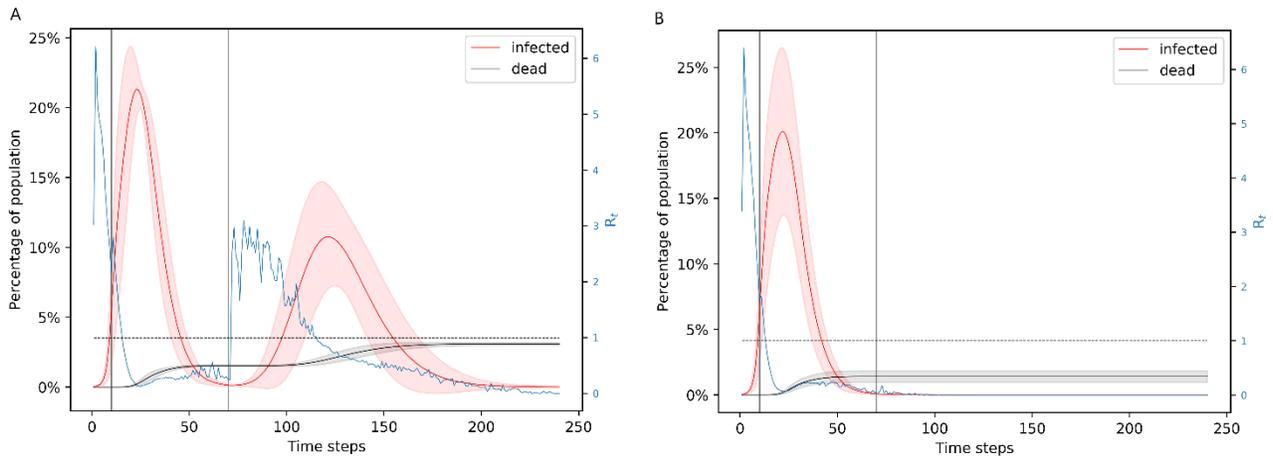

Figure 6A. Representation of the ~33% of simulations that have a second wave of infection for the day 10 to day 70 lockdown. B. Representation of the ~39% of simulations that have only one wave of infection for the day 10 to day 70 lockdown.

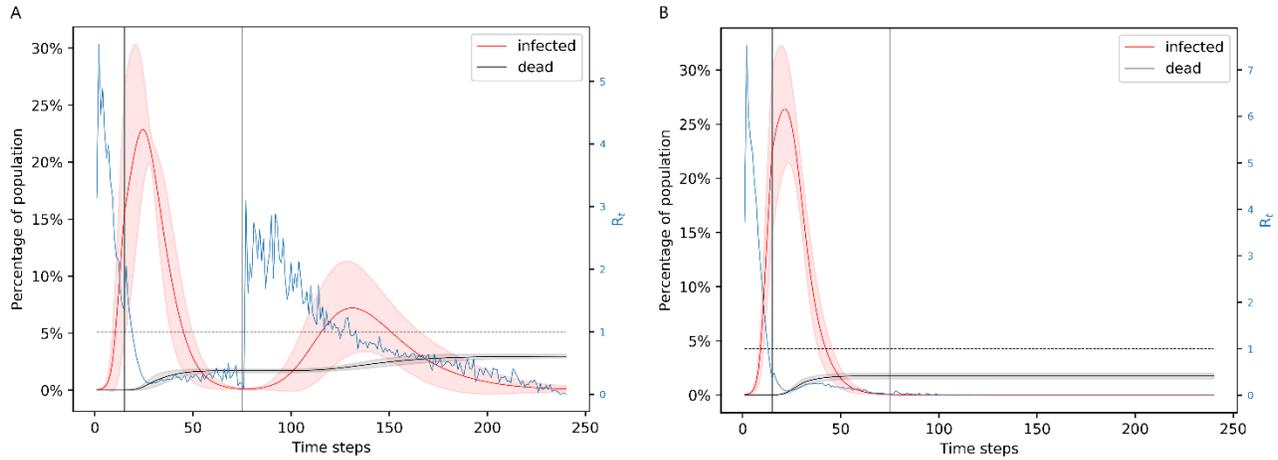

Figure 7A. Representation of the ~16% of simulations that have two waves of infection for the day 15 to day 75 lockdown. B. Representation of the ~58% of simulations that have only one wave of infection for the day 15 to day 75 lockdown.

### C. Phasing out of the lockdown

Here, we consider a slow phasing out of the lockdown. That is, facilities reopen slowly after the end of the lockdown and the population strictly adhere to social distancing, wearing masks and washing hands, etc. We implement this by restoring the network to its original state, that is, including all edges but reducing the transmission probability after the lockdown ends. The transmission probability then gradually increases to the original transmission probability in the window between the end of the lockdown and the day normalcy is resumed. We assume that the lockdown starts on day 5 and ends on day 65.

In the first case, shown in Figures 8A and 8B, the network follows a linearly increasing transmission probability from 0.01 on day 65 to 0.05 on day 80. This results in ~20% of the simulations having a second wave of infection, shown in Figure 8A. This second wave peaks around day 125 and infects nearly 10% of the population at the same time. The plot in this case (Fig 8A) is very similar to the plot in Figure 5A. However, the benefit of the gradual decrease of mitigation is reflected in much lower likelihood of the existence of a second wave (20%) compared to the likelihood when we resume to normalcy right after the end of the lockdown (40%). This case leads to an average of ~300 deaths. Figure 8B shows the simulation results when there is only one wave of infection; this case leads to an average of ~150 deaths. The gradual resumption to normalcy on day 80 results in a lower overall death count at ~200 compared to the ~250 with sudden resumption to normalcy (Figures 5A and 5B).

In the case shown in Figure 9A, the lockdown is from day 5 to day 65 and we resume to complete normalcy on day 95. The network follows a linearly increasing transmission rate from day 65 to day 95; on day 95, the transmission rate reaches the normal 5%. This case results in ~6% of simulations with a second wave of infection shown in Figure 9A. These simulations result in an average of ~300 deaths on average. The majority of the simulations have only one wave of infection and result in an average of ~150 deaths, shown in Figure 9B. This scenario results in ~160 deaths overall.

We also try the case where lockdown is from day 5 to day 65 and we resume to unrestricted interaction on day 110 (Figure 10). In this case, the probability of a second wave of infection is ~3% - this case is shown in Figure 10A. These simulations result in an average of ~300 deaths. The majority of the simulations, shown in Figure 10B have only one wave of infection and result in an average of ~150 deaths. Overall, this scenario results in ~150 deaths. Similar to the case presented in the previous subsection, the effective reproduction number for all of these cases has at least one peak with value ~6 occurring around day 3-4 and has another peak for the results that show a second wave of infection. The $R_t$ peak is around 20 days before the peak of the second wave of infection. With the much lower probability of a second wave of infection and fewer deaths, we can conclude that a 60-day lockdown with an extended period of gradual decrease of mitigation would be the best approach.

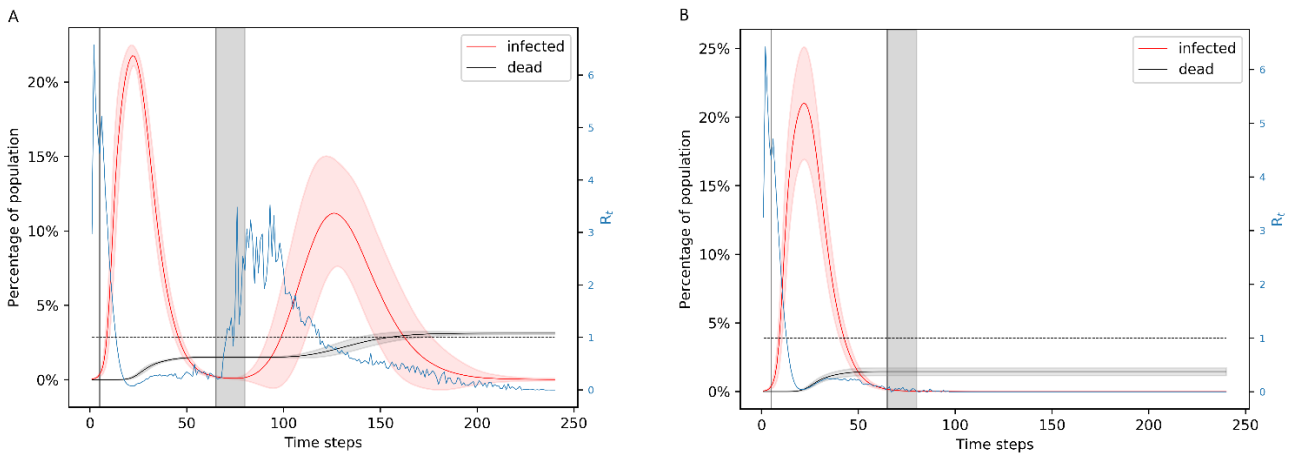

Figure 8A. Representation of the ~19% of simulation results that have a second wave of infection for the lockdown from day 5 to day 65 and complete normalcy resumes on day 80. The effective reproduction number starts to increase during the phasing out of the mitigation measures. B. Representation of the ~43% of simulation results that have only one wave of infection for the lockdown from day 5 to day 65 and complete normalcy resumes on day 80.

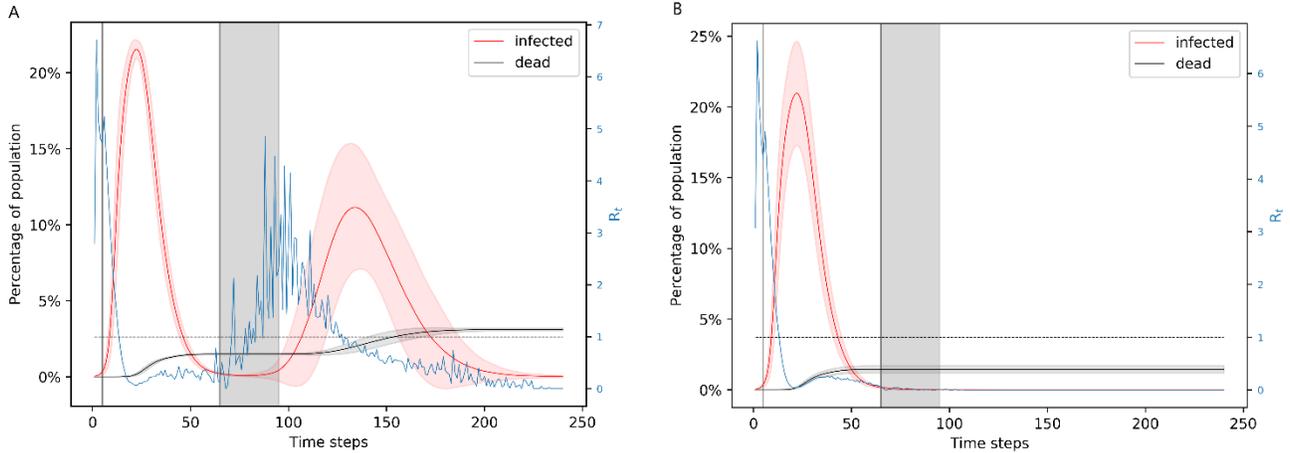

Figure 9A. Representation of the ~6% of simulation results that have a second wave of infection for the lockdown from day 5 to day 65 and complete normalcy resumes on day 95. B. Representation of the ~57% of simulation results that have only one wave of infection for the lockdown from day 5 to day 65 and complete normalcy resumes on day 95.

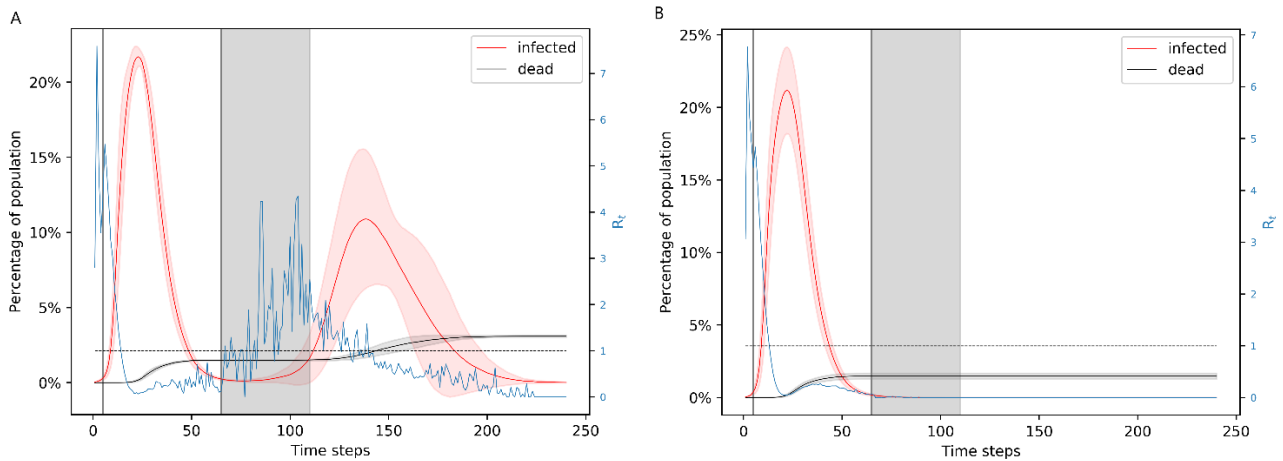

Figure 10A. Representation of the ~3% of simulation results that have a second wave of infection for the lockdown from day 5 to day 65 and complete normalcy resumes on day 110. B. Representation of the ~59% of simulation results that have only one wave of infection for the lockdown from day 5 to day 65 and complete normalcy resumes on day 110.

**IV.   Discussion**

We showed that a simple network model results in insightful modeling of the spread of an infectious disease like Covid-19 under social distancing conditions. The network generation method is based on parsimonious assumptions on the nature of everyday interaction and combine different walks of human interaction. We combine existing network models such as random graphs and scale-free networks, to construct an image of the real-world interaction network. The variable parameters in our network generation method allow for modeling different kinds of societies. We also elaborate on a simple disease spread mechanism in a network. Avoiding the complexity of more detailed spread processes that involve a latent period and pre-symptomatic phase, we employ few parameters to incorporate the most salient features of Covid-19 spread.

Our results show the trends of the probability of a second wave of infection and number of deaths for different lockdown situations. We explore different time-periods for lockdown starting with what is intuitively the safest option – a long lockdown window. We then explore shorter lockdown windows which give a moderately high probability for a second wave of infection. We explore different periods of gradual decrease of mitigation measures and find that a 60-day lockdown window paired with a 30-day or more of gradual increase of transmission leads to a very low probability of a second wave of infection and hence a lower death rate. Various modifications of this phasing out period could form promising directions for future work. For example, we can emulate the phasing out period by selectively adding edges to the sparser lockdown network instead of a varying transmission rate.

We also present the reproduction number of the disease calculated by case counting of the infection spread in the disease. There are various methods for estimating $R_t$ based on data-fitting and Monte Carlo simulations [15-18]. For example, one can fit the number of infectious individuals to a Poisson distribution and then use Markov Chain Monte Carlo to obtain a distribution of $R_t$ [18]. The estimates on the basic reproduction number of Covid-19 (prior to mitigation) are in the range 4 to 8; the estimated effective reproduction number reduces to ~1 or less than 1 after 30 days of lockdown [16,18,20]. In [20], for example, they show that in the District of Columbia the value of $R_t$ on March 17 was 8.19 and on April 1 it was 1.00. Thus, our simulation results for the basic and effective reproduction numbers are consistent with the literature. Our results show that the peak in the reproduction number foreshadows the peak in the number of infected nodes (see Figures 3-10). Given the foreshadowing of the infection peak by the $R_t$ peak, an adaptive mitigation strategy may be useful, namely monitoring the rise in the effective reproduction number and accordingly implementing social distancing regulations for the population.